\documentstyle[12pt,titlepage]{article}
\input epsfig.sty

\font\grande=cmr9.5 scaled \magstep4
\font\medio=cmr9.5 scaled \magstep2
\outer\def\beginsection#1\par{\medbreak\bigskip
      \message{#1}\leftline{\bf#1}\nobreak\medskip
\vskip-\parskip
      \noindent}

\def\laq{\raise 0.4ex\hbox{$<$}\kern -0.8em\lower 0.62
ex\hbox{$\sim$}}
\def\gaq{\raise 0.4ex\hbox{$>$}\kern -0.7em\lower 0.62
ex\hbox{$\sim$}}

\begin{document}
\bibliographystyle {unsrt}

\titlepage

\begin{flushright}
CERN-PH-TH/2006-253
\end{flushright}

\vspace{15mm}
\begin{center}
{\grande Gravitating multidefects from higher dimensions}\\
\vspace{20mm}
 Massimo Giovannini 
 \footnote{Electronic address: massimo.giovannini@cern.ch} \\
\vspace{6mm}

\vspace{0.3cm}
{{\sl Centro ``Enrico Fermi", Compendio del Viminale, Via 
Panisperna 89/A, 00184 Rome, Italy}}\\
\vspace{0.3cm}
{{\sl Department of Physics, Theory Division, CERN, 1211 Geneva 23, Switzerland}}
\vspace*{2cm}

\end{center}

\vskip 2cm
\centerline{\medio  Abstract}
Warped configurations admitting pairs of gravitating 
defects are analyzed. After devising a general method for the construction of multidefects, 
specific examples are presented in the case of higher-dimensional Einstein-Hilbert gravity.
 The obtained profiles describe diverse physical situations 
 such as  (topological) kink-antikink systems, pairs of non-topological 
 solitons and bound configurations  of a kink and of a non-topological 
 soliton. In all the mentioned cases the geometry is always well  behaved 
 (all relevant curvature invariants are regular) and tends 
to five-dimensional anti-de Sitter space-time for large asymptotic values 
of the bulk coordinate. Particular classes of solutions
can be generalized to the framework where the gravity part of the action includes, as a correction, the Euler-Gauss-Bonnet combination. 
After scrutinizing the structure of the zero modes, the obtained results are compared
with conventional gravitating configurations containing a single topological defect.

\noindent

\vspace{5mm}

\vfill
\newpage

\renewcommand{\theequation}{1.\arabic{equation}}
\section{Formulation of the problem}
\setcounter{equation}{0}

Solitons represent an important class of phenomena both in classical 
and quantum field theory. 
Solitonic solutions include monopoles, skyrmions, vortices an kinks 
\cite{RAJ1}.
It is well known that in $(1+1)$ dimensional field theories (static)
 topological solitons can arise 
 in combination with an appropriately non-linear interaction potential 
 \cite{RAJ2,MONT}. 
 These solutions are customarily called kinks and they 
 may arise either in the case of polynomial potential or in the case 
 of sine-Gordon potentials \cite{RAJ1,COL,RUB}.  In $(1+1)$ dimensions, spatial 
 infinity consists of two points (i.e. $\pm \infty$) and the kink solution 
 typically interpolates between two minima 
 of the underlying potential. The topological charge of the kink is positive.
 If the kink solution exist, there will also typically 
 exist the related antikink solution whose topological charge 
 will be opposite to the one of the kink. In recent years 
 general methods have been devised for the analysis of the complicated 
 nonlinear  problem arising  in the case of $(1+1)$-dimensional multidefect systems \cite{MOR,BAZ1,BAZ2,BAZ3,SOU}.
 
 It is also well known that, in $(1+1)$, dimensions non-topological 
 solitons can occur. These objects have vanishing topological 
 charge since the field profile vanishes for 
 large absolute value of the spatial coordinate \cite{TDLEE}. 
 
 It is finally known that, with the appropriate field content, multidefects
 do arise in $(1+1)$ dimensions \cite{RAJ2,MONT,BAZ2,SOU}. These solutions include, for 
 instance bound states of a kink and of an antikink and the so-called
 trapping bag solutions \cite{MONT,WEI} where a non-topological profile 
 is supplemented by a kink (or an antikink) profile. 
 
 There are mathematical and physical analogies between the defects of $(1+1)$ dimensional 
 field theories and the gravitating solitonic solutions of five-dimensional 
 gravity. In the latter case the r\^ole of the spatial coordinate is played 
 by the fifth dimension parametrizing the field profile in the bulk space-time.
 Along this perspective it is known that gravitating kink solutions 
 may arise, for instance, in the context of five-dimensional Einstein-Hilbert 
 gravity. Different groups analyzed {\em single} kink solutions 
 in higher-dimensional geometries. In particular examples are known both
 in flat five-dimensional space-time \cite{RUB1} (see also \cite{RUB2}),
and  in five-dimensional (warped) space-times 
\cite{TAM1,TAM2,GRE1,GRE2,MG1,MG2} (see also \cite{KK1,KK2} 
for an interesting perspective). While the compatibility 
of these configurations with higher-dimensional Einstein-Hilbert gravity 
has been ascertained, interesting generalizations contemplate the 
coupling of the scalar degree of freedom to the curvature 
\cite{TAM3,FAR1,FAR2}, the inclusion of Gauss-Bonnet 
self-interactions \cite{COR,MAVRO,MG4,MG5,GER}, the addition of more 
than one internal (warped) dimension  (see, for instance, \cite{COR2,COR3,MAXM}).
 While the features of the geometry may be diverse, it is 
 certainly plausible (even if not mandatory) that gravitating kink solutions lead to five-dimensional 
 anti-de Sitter geometry ( $\mathrm{AdS_{5}}$ in the following) for large absolute values 
 of the bulk coordinate.
In a related framework it is also possible to obtain kink profiles 
whose related geometry leads, for large absolute values 
of the bulk radius, to five-dimensional Minkowski space-time.
Moreover, single defects with non-topological features have been 
studied in \cite{MGNT}. It should be stressed that gravity is an essential 
ingredient for all the five-dimensional solitonic solutions 
mentioned in the present paragraph. By contrast, in $(1+1)$ dimensions, gravitational 
interactions are absent. This simple observation is the root of the physical differences 
occurring in the structure of the zero-modes of the system.

Indeed, gravitating solutions containing a single defect 
are rather intriguing since they may 
be used to localize fluctuations of various spin in five-dimensional 
gravity. They constitute a viable example of (static) brane models 
where the thickness of the brane does not vanish as it happens, instead 
in the Randall-Sundrum set-up \cite{RS1,RS2}.
The problem of localization of the fluctuations of single-defect 
models has been addressed in a number of ways. 
In \cite{MG1,MG2,MG3} a fully gauge-invariant formalism 
has been proposed and, subsequently, useful gauge-dependent 
approaches have also been developed \cite{RAN1,RAN2} (see also \cite{ZUL}). 

The problem to be discussed in the present paper can be formulated 
in analogy with the case of $(1+1)$-dimensional field theories. 
While we do know that multidefects are present in $(1+1)$ dimensions, 
it would be interesting to know if multidefects may arise 
in warped geometries. Consider, therefore, the simplest five-dimensional framework 
allowing for gravitating multidefects and described by the action \footnote{Latin 
(uppercase) letters run over the higher-dimensional space-time while 
Greek (lowercase) indices are defined over the four-dimensional space-time.}
\begin{equation}
S_{\mathrm{EH}}= \int d^{5} x \sqrt{|G|} \biggl[ - \frac{R}{2\kappa} + 
\frac{1}{2} G^{A B} \partial_{A} \varphi \partial_{B} \varphi + 
\frac{1}{2} G^{AB} \partial_{A} \chi \partial_{B} \chi - W(\varphi,\chi)\biggr],
\label{action}
\end{equation}
where $\kappa = 8\pi/M_{5}^3$, $G_{AB}$ is the five-dimensional 
metric tensor and $R$ the five-dimensional Ricci scalar. In Eq. (\ref{action})
$\phi$ and $\chi$ are two scalar degrees of freedom and $W(\phi,\chi)$ 
is the potential containing both the self-interactions of $\phi$ and $\chi$ 
as well as their mutual interactions.
It will be shown that solutions containing pairs of gravitating defects  
are a common feature of the theory defined by Eq. (\ref{action})
 and can be obtained with an appropriate (but rather general) constructive
 procedure. The discussion presented here is limited to pair of defects, however, it seems 
 rather plausible to extend the discussion to an even larger number of fields.

It was recently pointed out that kink-antikink solutions as well as 
trapping bag solutions may arise in the presence of suitably 
tuned Gauss-Bonnet corrections to the Einstein-Hilbert action \cite{MG6}.
In the present analysis those considerations will be extended and it will 
be shown, in particular, that Gauss-Bonnet corrections are a useful 
addition but they are not crucial for the existence of gravitating multidefects. 
After discussing the physical properties of the new solutions 
containing pairs of defects it will be also interesting to 
develop the gauge-invariant approach proposed in \cite{MG1,MG2,MG3}
to include the case of configurations containing more than one defect.  
 
The plan of the present paper is then the following. In sec. 2 
the framework of the present analysis will be 
explained. In sec. 3, after developing a rather general method for the 
integration of the system,  kink-antikink 
configurations  will be specifically analyzed.
Section 4 deals primarily with bound systems of topological and non-topological solitons. The analysis of the zero modes 
of the system is contained in sec. 5. Section 6 contains the concluding remarks and a summary of the main findings of the present investigation.

\renewcommand{\theequation}{2.\arabic{equation}}
\section{The general framework}
\setcounter{equation}{0}
Consider first the case of five-dimensional Einstein-Hilbert gravity 
characterized by the action (\ref{action}). The field equations 
derived from Eq. (\ref{action}) simply become:
\begin{eqnarray}
&& R_{AB} - \frac{1}{2} G_{AB} R = \kappa T_{A B},
\label{EH1}\\
&& G^{A B} \nabla_{A} \nabla_{B} \phi + \frac{\partial W}{\partial \phi} =0,
\label{EH2}\\
&& G^{A B} \nabla_{A} \nabla_{B} \chi + \frac{\partial W}{\partial \chi} =0,
\label{EH3}
\end{eqnarray}
where $R_{AB}$ is the Ricci tensor and $\nabla_{A}$ is the covariant 
derivative constructed from the five-dimensional metric $G_{AB}$.
The energy-momentum tensor of the system, $T_{AB}$, 
 is given, according to Eq. (\ref{action}) by 
\begin{equation}
T_{AB} = \partial_{A} \phi \partial_{B} \phi + \partial_{A} \chi \partial_{B}\chi  - 
G_{AB} \biggl[ \frac{G^{MN}}{2} \partial_{M} \phi \partial_{N} \phi + 
\frac{G^{M N}}{2} \partial_{M} \chi \partial_{N} \chi - W(\phi,\chi)\biggr].
\label{TAB}
\end{equation} 
Consider now, in particular, the five-dimensional (warped) line element
\begin{equation}
ds^2 = a^2(w) [ \eta_{\mu\nu} dx^{\mu} dx^{\nu} - dw^2],
\label{met1}
\end{equation}
where $w$ is the bulk coordinate, $a(w)$ the warp factor and $\eta_{\mu\nu}$ the 
four-dimensional Minkowski metric.
In the case described by Eq. (\ref{met1}), Eqs. (\ref{EH1}), (\ref{EH2}) and (\ref{EH3}) 
become, respectively, 
\begin{eqnarray}
&& {\mathcal H}' + {\mathcal H}^2 = - \frac{\kappa}{3} \biggl[ \frac{{\phi'}^2}{2} +  \frac{{\chi'}^2}{2} + W a^2\biggr],
\label{EHex1}\\
&& {\mathcal H}^2 = \frac{\kappa}{6}  \biggl[ \frac{{\phi'}^2}{2} +  \frac{{\chi'}^2}{2} - W a^2\biggr],
\label{EHex2}\\
&& \phi'' + 3 {\mathcal H} \phi' - a^2 \frac{\partial W}{\partial \phi} =0,
\label{EHex3}\\
&& \chi'' + 3 {\mathcal H} \chi' - a^2 \frac{\partial W}{\partial \chi} =0,
\label{EHex4}
\end{eqnarray}
where the prime denotes a derivation with respect to the bulk coordinate 
$w$ and, within this notation, ${\mathcal H} = (\ln{a})'$.
By linearly combining Eqs. (\ref{EHex1}) and (\ref{EHex2})  the following 
useful pair of equations can be readily obtained:
\begin{eqnarray}
&&{\phi'}^2 + {\chi'}^2 = \frac{3}{\kappa} ( {\mathcal H}^2 - {\mathcal H}'),
\label{EHexa}\\
&& a^2  W = - \frac{3}{2 \kappa} ({\mathcal H}' + 3 {\mathcal H}^2).
\label{EHexb}
\end{eqnarray}
 Needless to say that Eqs. (\ref{EHexa}) and (\ref{EHexb}) are fully 
 equivalent to Eqs. (\ref{EHex1}) and (\ref{EHex2}).
In the presence of more $3$ transverse dimensions, the line element 
(\ref{met1}) can be generalized as 
\begin{equation}
ds^2 = a^2(w) [ dt^2 - dx_{1}^2 - dx_{2}^2 -...- d x_{ d}^2 - dw^2],
\label{met2}
\end{equation}
where the ellipses stand for the $d$ (transverse) spatial coordinates  
while $w$ still denotes the bulk coordinate. The overall 
dimensionality of the space-time $D$ will then be given by $D = d + 2$.
Consequently, in the case of the line element of Eq. (\ref{met2}), the analog 
of Eqs. (\ref{EHexa}) and (\ref{EHexb}) will be 
\begin{eqnarray}
&&{\phi'}^2 + {\chi'}^2 = \frac{d}{\kappa} ( {\mathcal H}^2 - {\mathcal H}'),
\label{EHexc}\\
&& W(\phi,\chi) = - \frac{d}{2 \kappa a^2} ( {\mathcal H}' + d {\mathcal H}^2).
\label{EHexd}
\end{eqnarray}

Another type of generalization of the system under 
discussion concerns the addition of quadratic corrections 
to the Einstein-Hilbert term. In this case the total action is given by 
\begin{equation}
S = S_{\mathrm{EH}} + S_{\mathrm{GB}},
\label{sum}
\end{equation}
where 
\begin{equation}
S_{\mathrm GB}
 = - \alpha'  \int d^{5} x \sqrt{|G|} {\mathcal R}_{\mathrm{GB}}^2,
\label{GBaction}
\end{equation}
is the Gauss-Bonnet action\footnote{In Eq. (\ref{GBaction}) $\alpha'$ is a constant 
with dimensions, in natural gravitational units $2\kappa =1$, of an inverse length. It is practical, at the 
level of the Einstein-Lanczos equations, to define an effective coupling $\overline{\alpha} = 2\kappa \alpha'$ (see, below, Eqs. (\ref{EL1}) and (\ref{EL2}).} 
written in terms of the Euler-Gauss-Bonnet combination
\footnote{ The Euler-Gauss-Bonnet (or simply Gauss-Bonnet for short) combination 
\cite{lan,lov,mad1,mad2,zw,des,ts,cal,sen}  
arises in different higher-dimensional theories and it does 
also appear in the low-energy string effective action as first 
correction in the string tension expansion.}, i.e. 
\begin{equation}
{\mathcal R}^2_{\mathrm{GB}} = R_{ABCD}R^{ABCD} - 4 R_{AB} R^{AB} + R^2.
\label{GBcombination}
\end{equation}
In four space-time dimensions, the Gauss-Bonnet combination (\ref{GBcombination}) 
is a topological term \cite{lan,lov} and it coincides with the Euler invariant. This observation 
implies that, in four space-time dimensions, the contribution of the four-dimensional analog 
of Eq. (\ref{GBaction}) to the equations of motion can be rearranged in a four-divergence \cite{lan}. 

In more than four space-time dimensions, the contribution of the Gauss-Bonnet 
combination leads to a ghost-free theory \cite{zw}.
 In particular, the variation of the Gauss-Bonnet action (\ref{GBaction}) 
brings, at the right hand side of Eq. (\ref{EH1}), a new term which can be 
written as $ 2 \kappa \alpha' {\mathcal Q}_{AB}$ 
where
\begin{equation}
{\mathcal Q}_{A}^{B} = \frac{1}{2} \delta_{A}^{B} {\mathcal R}^2_{\mathrm{GB}} -
2 R\, R_{A}^{B} + 4 R_{A C}R^{CB} + 4 R_{CD} R_{A}^{~~CBD} - 2 R_{ACDE}R^{BCDE}.
\label{lanczos}
\end{equation}
is the so-called Lanczos  tensor \cite{lan,lov,mad1,mad2} (see also \cite{mad3} 
for an interesting review on Gauss-Bonnet gravity and Einstein-Lanczos equations).

The relevant featture to be appreciated is that the inclusion of the contribution 
provided by Eq. (\ref{lanczos}) generalizes Eqs. (\ref{EHexa}) and (\ref{EHexb}) to a form which is, however, still tractable:
\begin{eqnarray}
&& {\phi'}^2 + {\chi'}^2 = \frac{3}{\kappa} ( {\mathcal H}^2 - {\mathcal  H}')\biggl[ 1 - \frac{4\overline{\alpha}}{a^2} {\mathcal H}^2 \biggr],
\label{EL1}\\
&& W(\phi,\chi) = - \frac{3}{2 \kappa a^2} \biggl\{ ({\mathcal H}^2  - {\mathcal H}') + 2 
({\mathcal H}^2  + {\mathcal H}') \biggl[ 1 - \frac{2 \overline{\alpha}}{a^2} {\mathcal H}^2\biggr]\biggr\},
\label{EL2}
\end{eqnarray}
where $\overline{\alpha} = 2 \kappa \alpha'$. Note that $ \overline{\alpha}$ has dimensions
of a squared length. Equations (\ref{EL1}) and (\ref{EL2}) supplemented 
by Eqs. (\ref{EHex3}) and (\ref{EHex4}) (which are not affected by the quadratic 
corrections) form a closed set 
of equations which will be denoted, for short, as the 
Einstein-lanczos system.

Finally, as in the case of Einstein-Hilbert gravity, the Einstein-Lanczos equations (\ref{EL1}) and (\ref{EL2}) 
can be further generalized to include $d$ transverse dimensions. 
In fact Eq. (\ref{met2}) implies that Eqs. 
(\ref{EL1}) and (\ref{EL2}) are generalized to:
\begin{eqnarray}
&& {\phi'}^2 + {\chi'}^2 = \frac{d}{\kappa}( {\mathcal H}^2 - {\mathcal  H}')
 \biggl[ 1 - \frac{2\overline{\alpha}( d-1) (d -2)}{a^2} {\mathcal H}^2 \biggr],
\label{EL3}\\
&& W(\phi,\chi) = - \frac{d}{2 \kappa a^2} \biggl\{ ({\mathcal H}^2  - {\mathcal H}') +  
[(d -1){\mathcal H}^2  + 2 {\mathcal H}'] 
\biggl[ 1 - \frac{ \overline{\alpha} (d -1) ( d-2)}{a^2} {\mathcal H}^2\biggr]\biggr\},
\label{EL4}
\end{eqnarray}

In the limit when one of the two scalar degrees of freedom is absent 
it is known that gravitating defects are present and may even lead 
to realistic warped geometries with well defined $\mathrm{AdS}_{5}$
limit for large value of the bulk coordinate. For instance 
in \cite{TAM1,TAM2} and in \cite{GRE1,GRE2,MG1,MG2} 
kink solutions have been presented in the context of Einstein-Hilbert
gravity and in the presence of either polynomial or generalized 
sine-gordon potentials. Single kink solutions may also be obtained
in Gauss-Bonnet gravity \cite{COR,MAVRO,MG4}. 
The aim of the subsequent sections will be to show 
that bound systems of two gravitating defects may also arise 
naturally in the framework of the present section. Furthermore, particular 
attention will be given to the case where the five-dimensional geometry 
possesses the  desired  $\mathrm{AdS}_{5}$ limit.

\renewcommand{\theequation}{3.\arabic{equation}}
\section{Soliton-antisoliton systems}
\setcounter{equation}{0}

Consider, to begin, the case of five-dimensional Einstein-Hilbert gravity with 
one non-compact extra-dimension denoted by $w$. In this case 
a general ansatz for the soliton-antisoliton system can be written as 
\begin{equation}
\phi(w) = v \sqrt{ 1 + g(w)},\qquad
\chi(w) = v \sqrt{ 1 - g(w)},
\label{santis1}
\end{equation}
where $v$ is a dimension-full constant (i.e. $[v] = L^{-3/2}$) and $g(w)$ is a dimensionless 
function of the bulk radius satisfying the following set of properties:
\begin{itemize}
\item{} $g(w)$ is a continuous and differentiable function of $w$;
\item{} $g(w)$ is a monotonic function of $w$;
\item{} the derivatives of $g(w)$ are also continuous at least 
up to the second derivative (i.e. $g'(w)$ and $g''(w)$ are continuous functions of $w$).
\end{itemize}
Since the ansatz (\ref{santis1}) must describe a bound system of a kink and of an antikink it must also be required that 
\begin{equation}
\lim_{w\to \pm \infty} g(w) = \pm 1.
\label{limit}
\end{equation}
The numerical value of the limit is just a reflection of the parametrization (\ref{santis1}) but it is important, for more general parametrizations,  that $g(w)$ goes to a constant for large values of the bulk coordinate.
Finally, it is obvious that since $g(w)$ is monotonic, $g'(w)$ will never vanish 
for any finite value of the bulk radius. 

Inserting Eq. (\ref{santis1}) into Eqs. (\ref{EHexa}) and (\ref{EHexb}) a necessary condition on the 
functional form of $g'(w)$ can be obtained:
\begin{equation}
\biggl( \frac{d g}{dw}\biggr)^2 = \frac{6}{\kappa v^2} ( 1 - g^2 ) ({\mathcal H}^2 - {\mathcal H}').
\label{gcond}
\end{equation}
Once the geometry is specified, Eq. (\ref{gcond}) allows to obtain the explicit form of $g(w)$. Then, by fixing the integration constants it is possible to satisfy the conditions required for the existence of a kink-antikink solution. 

Consider then, as an example, 
 the simple case when the warp factor is given by 
\begin{equation}
a(w) = \frac{1}{\sqrt{x^2 + 1}}, \qquad x = b\, w,
\label{ADS5}
\end{equation}
where $b$ is a dimensionfull constant which will eventually fix the thickness 
of the configuration. Using Eq. (\ref{ADS5}) into Eq. (\ref{gcond}) the following differential equation can be readily obtained 
\begin{equation}
\frac{d g}{d x} = \pm \sqrt{\frac{6}{\kappa v^2}} \frac{\sqrt{ 1 - g^2}}{x^2 + 1},
\label{gcond1}
\end{equation}
whose solution is 
\begin{equation}
g(x) = \pm \frac{x}{\sqrt{x^2 + 1}}, \qquad k v^2 = 6.
\label{gcond2}
\end{equation}
The plus sign at the right hand side of Eq. (\ref{gcond2}) will be conventionally chosen. This choice is, indeed, rather general and it implies that, according to Eq. (\ref{santis1}), 
$\phi(w)$ describes a kink while $\chi(w)$ describes and antikink. In fact, defining the 
topological charges as 
\begin{equation}
Q_{\phi} = \frac{1}{2\pi} \int_{-\infty}^{+\infty} \frac{\partial \phi}{\partial w} dw, \qquad 
Q_{\chi} = \frac{1}{2\pi} \int_{-\infty}^{+\infty} \frac{\partial \chi}{\partial w} dw,
\label{topcha}
\end{equation}
we will have, according to Eqs. (\ref{santis1}) and (\ref{gcond2}), that $\pi Q_{\phi} = v/\sqrt{2} >0 $
and that $\pi Q_{\chi} = - v/\sqrt{2} < 0$. If $g(x) \to - g(x)$, the transformed ansatz is still a solution 
of Eq. (\ref{gcond}) (corresponding to the lower sign in Eq. (\ref{gcond2})) and it implies that 
$\phi \to \chi$ and that $\chi \to \phi$. In other words, if $g\to -g$ the kink turns into an antikink and viceversa.
By choosing the upper sign in Eq. (\ref{gcond2}) we can also write the differential relation appearing in Eq. (\ref{gcond1}) as 
\begin{equation}
\frac{d g}{d x} = ( 1 - g^2)^{3/2}.
\label{gcond3}
\end{equation}
Equations (\ref{EHex3}), (\ref{EHex4}) and (\ref{EHexb}) allow then to compute the specific form 
of the potential first as a function of $g$ and then as a function of $\phi$ and $\chi$. 
Using Eqs. (\ref{ADS5}), (\ref{gcond2}) and (\ref{gcond3}) and recalling Eq. (\ref{santis1}) 
we will have that, as a function of $g$
\begin{eqnarray}
&& W(g) = \frac{3 b^2}{2 \kappa} ( 1 - 5 g^2),
\label{potg1}\\
&& \frac{\partial W}{\partial\phi}(g) = - \frac{b^2}{4} v \sqrt{ 1 + g} ( 1 - g) ( 1 + 11 g),
\label{potg2}\\
&& \frac{\partial W}{\partial\chi}(g) =- \frac{b^2}{4} v \sqrt{ 1 - g} ( 1 + g) ( 1 - 11 g).
\label{potg3}
\end{eqnarray}
Equations (\ref{potg1}), (\ref{potg2}) and (\ref{potg3}) allow to determine the explicit form 
of the potential which is 
\begin{eqnarray}
W(\phi,\chi) &=& \frac{9 b^2}{2 \kappa v^2}( \phi^2 + \chi^2) - \frac{15 b^2}{4 \kappa v^4} ( \phi^4 + \chi^4) 
\nonumber\\
&+& (\phi^2 + \chi^2 - 2 v^2) \biggl[ \frac{11 b^2}{16 v^4} ( \phi^4 + \chi^4) - \frac{b^2}{2 v^2} (\phi^2 + \chi^2)\biggr].
\label{pot1ex}
\end{eqnarray}
It can be explicitly verified, as a useful cross-check, that Eq. (\ref{pot1ex}), once inserted 
into Eqs. (\ref{EHex1})--(\ref{EHex4}), solves the system provided the warp factor is given by Eq. (\ref{ADS5}) and 
the field profiles are the ones of Eq. (\ref{santis1}).

The obtained result can be generalized to the case of $d$ transverse dimensions (i.e. 
Eqs. (\ref{met2})). The calculation follows, in this case, exactly the same steps 
discussed for the previous five-dimensional example. 
Less obvious, even if solvable, is the generalization of the obtained results to the situation where 
the Gauss-Bonnet self-interactions are present.  

Let us  show, as a premise, that the ansatz of Eq. (\ref{santis1}) is {\em not} a solution of the five-dimensional Einstein-Lanczos 
system, i.e. Eqs. (\ref{EL1}) and (\ref{EL2}) supplemented by Eqs. (\ref{EHex3}) and (\ref{EHex4}).
Equation (\ref{ADS5}) can be generalized to the following 
form
\begin{equation}
a(w) = \frac{a_0}{\sqrt{x^2 + 1}},
\label{ADS5GB}
\end{equation}
with $a_{0} = 2 \sqrt{\overline{\alpha}} b$, as required by the compatibility of Eq. (\ref{ADS5GB}) with Eq. (\ref{EL1}).
Thus, inserting Eq. (\ref{ADS5}) and Eq. (\ref{santis1}) into Eq. (\ref{EL1}) the following 
differential condition on $g$ can be obtained:
\begin{equation}
\biggl( \frac{d g}{dx}\biggr)^2 = \beta^2 \frac{1 - g^2}{(1 + x^2)^3},\qquad \beta^2 = \frac{6}{ \kappa v^2}. 
\label{gcond4}
\end{equation}
Now, in the case $\beta = \pi/2$ the solution of Eq. (\ref{gcond4})  is 
\begin{equation}
g(x) = \sin{\biggl( \frac{\pi}{2} \frac{x}{\sqrt{x^2 + 1}}\biggr)}.
\label{gcond5}
\end{equation}
Equation (\ref{gcond5}) certainly satisfies the conditions required for the existence of a 
kink-antikink solution. However, the compatibility with Eq. (\ref{EL2}) as well as with 
Eqs. (\ref{EHex1}) and (\ref{EHex2}) cannot be achieved. 

It can be actually shown \cite{MG6} that the correct ansatz leading to a consistent solution 
of the EInstein-Lanczos system is given, in this case, by the following expressions
\begin{eqnarray}
&& \phi(w) = \frac{v}{\sqrt{2}} \biggl(1 + \frac{x}{\sqrt{x^2 + 1}}\biggr)^{3/2},
\label{phiGB}\\
&& \chi(w) = \frac{v}{\sqrt{2}} \biggl(1 - \frac{x}{\sqrt{x^2 + 1}}\biggr)^{3/2},
\label{chiGB}
\end{eqnarray}
having chosen the warp factor in the form (\ref{ADS5GB}) with $a_{0} = 2 \sqrt{\overline{\alpha}} b$ and with 
$\kappa v^2 = 4/3$. In this case the potential $W(\phi,\chi)$ can be written as 
\begin{eqnarray}
W(\tilde{\phi},\tilde{\chi}) &=& \frac{3 v^2}{2 \overline{\alpha}} ( \tilde{\phi}^2 + \tilde{\chi}^2)^2 - 
\frac{3 v^2}{\overline{\alpha}} (\tilde{\phi}^2 + \tilde{\chi}^2) + 
\frac{15}{16} \frac{v^2}{\overline{\alpha}} 
\nonumber\\
&+& 
\frac{7}{2} \frac{v^2}{\overline{\alpha}}( |\tilde{\phi}|^{2/3} + |\tilde{\chi}|^{2/3} -1) ( 1 - \tilde{\phi}^2 - \tilde{\chi}^2)^2,
\label{potkkGB}
\end{eqnarray}
where, for notational convenience, the canonical fields $\tilde{\phi}$ and $\tilde{\chi}$ have been 
introduced (note that $\phi = 2\,v\,\tilde{\phi}$ and $\chi = 2 \,v\, \tilde{\chi}$).
The solution expressed by Eqs. (\ref{phiGB}) and (\ref{chiGB}) is qualitatively similar to the one 
of Eq. (\ref{santis1}) but it is, at the same time, mathematically different.
Even if the geometry is given, in both cases, by a warp factor that tends to $\mathrm{AdS_{5}}$ for 
large absolute value of the bulk coordinate, the $\mathrm{AdS_{5}}$ radius is different in the two cases given, respectively,
by Eqs. (\ref{ADS5}) and (\ref{ADS5GB}) the difference being given by $a_{0} = 2 \sqrt{\overline{\alpha}} b$.
Finally, the solution (\ref{phiGB}) and (\ref{chiGB}) holds for $\phi >0$ and $\chi>0$. However, 
also $\phi \to -\phi$ and $\chi\to -\chi$ is a solution of the system as it is clear from the absolute values appearing in Eq. (\ref{potkkGB}).

\begin{figure}
\epsfxsize = 11 cm
\centerline{\epsffile{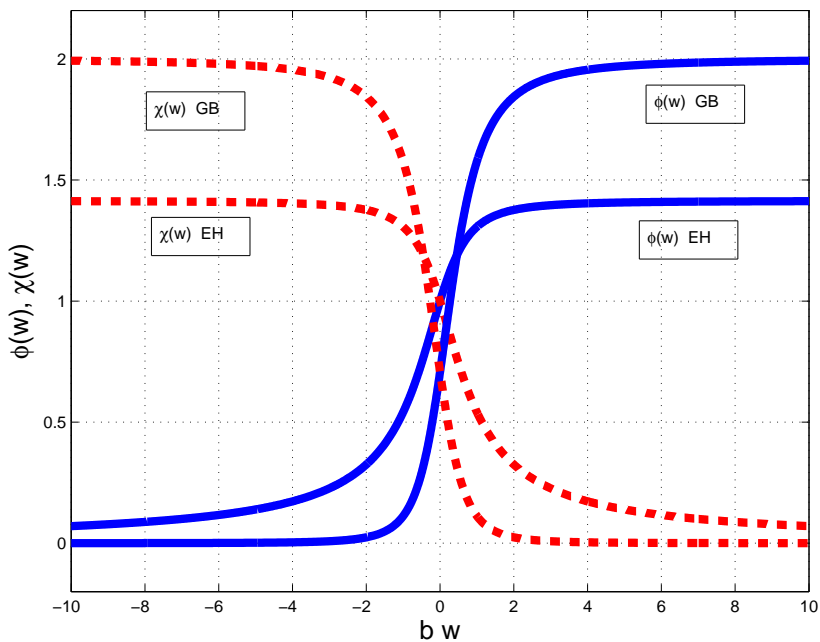}}
\caption[a]{The Einstein-Hilbert (EH) and the Gauss-Bonnet (GB)
kink-antikink systems are illustrated, by fixing, conventionally $v =1$ in natural 
gravitational units $2\kappa =1$. The full lines denote the kink profiles while 
the dashed lines denote the antikink profiles.}
\label{F1}
\end{figure}

In Fig. \ref{F1} the kink-antikink solutions are illustrated 
for the cases of the Einstein-Hilbert kink (see Eqs. (\ref{santis1}) 
and (\ref{gcond2})) and of the Gauss-Bonnet kink (see 
Eqs. (\ref{phiGB}) and (\ref{chiGB})). 
In Fig. \ref{F1} the dashed line denotes the antikink while 
the full line illustrates the kink. 
It should be stressed that the main objective of this paper is not to find 
single gravitating kinks (or single antikinks) but rather to find solutions 
where two field configurations are simultaneously present with 
two opposite topological charges.

While the solutions reported in Fig. \ref{F1} are interesting 
by themselves, it is worth stressing that, in the present section, a 
rather general method for constructing multidefects solutions 
has been proposed. The same method will also be exploited in the forthcoming sections in order to generalize the present considerations 
to the case of bound systems of topological and non-topological
solitons.

As a final remark, it should be stressed that the solutions discussed in the present section 
can be generalized to the case where there are, generically, $d$ transverse dimensions.
In fact, owing to the form of Eqs. (\ref{EHexc})--(\ref{EHexd}) and of Eqs. (\ref{EL3})--(\ref{EL4}),
the same analytical steps can be successfully carried on.
\renewcommand{\theequation}{4.\arabic{equation}}
\section{Topological and non-topological solitons}
\setcounter{equation}{0}

In the present section a general class of solutions containing pairs of defects 
will be presented. The new feature of this class is that, for some values of a discrete parameter, the 
solutions describe a system containing simultaneously, a topological and a non-topological 
defect. In the same class of solutions, it is also possible to find configurations 
that can be interpreted as a classically bound system of two non-topological profiles.

Consider, for this purpose, the following form of the warp factor:
\begin{equation}
a(w) = [ x^{2\nu} + 1]^{- \frac{1}{2\nu}}, \qquad x = b w,
\label{wfnu}
\end{equation}
where $\nu$ is a natural number. Following a similar procedure of the 
one presented in the  previous section consider 
also the following ansatz for the field profiles:
\begin{eqnarray}
&& \phi(w) = v \{ [ 1 + h(w)]^{3/2} + [ 1- h(w)]^{3/2}\}, 
\label{phNT}\\
&& \chi(w) = v \{ [ 1 + h(w)]^{3/2} - [ 1- h(w)]^{3/2}\},
\label{chNT}
\end{eqnarray}
where $h(w)$ is a continuous function sharing, for a topological soliton, exactly the same properties 
of $g(w)$.  In particular, for a topological soliton, it will be required that 
\begin{equation}
\lim_{w\to \pm \infty} h(w) = \pm 1.
\label{hlimit}
\end{equation} 
The analog of Eq. (\ref{hlimit}) in the case of a non-topological profile 
will instead be:
\begin{equation}
\lim_{|w|\to \infty} h(w) = \pm 1.
\label{hlimit2}
\end{equation} 
In spite of a superficial similarity, Eqs. (\ref{hlimit}) 
and (\ref{hlimit2}) imply, indeed, different boundary conditions. In the case of Eq. (\ref{hlimit}) 
the function $h(w)$ goes either to $-1$ or to $+1$ provided $w$ goes, respectively, 
either to $- \infty$ or to $+\infty$. For the boundary conditions expressed by 
Eq. (\ref{hlimit2}), the function $h(w)$ goes to the same value as soon as $|w| \to \infty$. This value can be either $+1$ or $-1$ since, as it will be discussed 
in a moment, the change $h \to -h$ still leads to a viable solution. It is important to remark that, as 
a consequence off the different boundary conditions at infinity, the function $h(w)$ is either 
monotonic (topological profile) or non-monotonic (non-topological profile). In this 
sense $h(w)$ shares exactly the same properties of $g(w)$ {\em only} in the case of topological 
configurations.

Working in the framework of five-dimensional Einstein-Hilbert gravity, Eq. (\ref{EHexa}) 
allows the determination of $h(w)$ whose functional form must obey 
the following differential condition
\begin{equation}
\biggl(\frac{d h}{d x}\biggr)^2 = \biggl(\frac{2\nu -1}{3 \kappa v^2}\biggr) \frac{x^{2\nu - 2}}{x^{2\nu} + 1},
\label{hcond1}
\end{equation}
whose solution is 
\begin{equation}
h(x) = \frac{2}{\pi}  \arctan{(x^{\nu})}, \qquad 
\kappa v^2 = \frac{\pi^2}{12} \frac{2 \nu -1 }{\nu^2}.
\label{hcond2}
\end{equation}
Equation (\ref{hcond2}) implies that if $\nu$ is an odd integer, $h(x)$ will necessarily 
interpolate between $-1$ and $+1$ (as required in the first set of boundary 
conditions written in Eq. (\ref{hlimit})). If, on the contrary,  $\nu$ is an even integer, $h(x)$ will 
always go to $+1$ as soon as $|x| \to \infty$ (as required in the second set 
of boundary conditions written in Eq. (\ref{hlimit2})). Furthermore, in the case 
of even-$\nu$, $h(x)$ will have a global minimum in $x=0$. 
As already anticipated in the previous paragraph, the form of the differential condition (\ref{hcond1}) 
implies that if $h(x)$ is a solution, also $-h(x)$  will be a solution. If $h\to - h$, Eqs. 
(\ref{phNT}) and (\ref{chNT}) imply that $ \phi \to \phi$ and $\chi\to -\chi$.

The difference between the solutions with odd-$\nu$ and the ones with even-$\nu$ 
implies also different physical properties which will be one of the themes of the forthcoming discussion.
The potential satisfying Eqs. (\ref{EHex3}), (\ref{EHex4}) and (\ref{EHexb}) (or, equivalently, Eqs. 
(\ref{EHex1})--(\ref{EHex4})) can be written as 
\begin{equation}
W(\phi,\chi) = \frac{3 b^2}{2 \kappa} [\sin^2{\sigma}]^{\frac{\nu -1}{\nu}} 
[ ( 2 \nu -1) - (2 \nu + 3) \sin^2{\sigma}] + [|\tilde{\phi} + \tilde{\chi}|^{2/3} + |\tilde{\phi} - \tilde{\chi}|^{2/3} -2] {\mathcal A},
\label{POTnu}
\end{equation}
where, as usual, $\phi = 2 v \tilde{\phi}$ and $\chi = 2 v \tilde{\chi}$. The functions 
$\sigma(\tilde{\phi},\tilde{\chi})$ and ${\mathcal A}(\tilde{\phi},\tilde{\chi})$ appearing in Eq. (\ref{POTnu}) 
are defined, respectively, as:
\begin{eqnarray}
&&{\mathcal A}(\tilde{\phi}, \tilde{\chi}) = \frac{18 b^2\, v^2 \,\nu}{\pi^2} \cos{\sigma}
 \{ \sigma [\sin^2{\sigma}]^{\frac{\nu-2}{2\nu}} [ (\nu -1) - ( 2 \nu + 3) \sin^2{\sigma}] 
\nonumber\\ 
&&+ \frac{\nu}{2} \cos{\sigma} [\sin^2{\sigma}]^{\frac{\nu -1}{\nu}}\},
\label{Anu}\\
&&\sigma(\tilde{\phi},\tilde{\chi}) = \frac{\pi}{4} [ |\tilde{\phi} + \tilde{\chi}|^{2/3} - |\tilde{\phi} - \tilde{\chi}|^{2/3}].
\label{sigmanu}
\end{eqnarray}

Let us now scrutinize the physical properties of the obtained solution.
From Eq. (\ref{wfnu}), (both for even and odd values of $\nu$), the warp factor 
always tend to $\mathrm{AdS_{5}}$ for $|w| \to \infty$ i.e. 
\begin{equation}
\lim_{w \to \pm \infty} a(w) = \frac{1}{b|w|}.
\label{limita}
\end{equation}
The limit (\ref{limita}) reproduces the asymptotic behaviour of the 
warp factors considered in Eqs. (\ref{ADS5}) and (\ref{ADS5GB}).
Furthermore, in the cases of Eqs. (\ref{ADS5}), (\ref{ADS5GB}) and 
(\ref{wfnu}) the geometry is always regular since all the 
curvature invariants are never divergent at finite values of the 
bulk coordinate and they go to a constant in the limit $|w|\to \infty$ 
as implied by the $\mathrm{AdS}_{5}$ limit of the warp factors.

As previously argued in connection with the properties 
of $h(x)$ the features of the solutions differ depending 
on the even or odd values of $\nu$. 
Consider the first few odd values of $\nu$, i.e. $\nu = 1, 3, 5...\,\,$. The analysis 
of these situations will allow to gain intuition on the  more general case.  
\begin{figure}
\begin{center}
\begin{tabular}{|c|c|}
      \hline
      \hbox{\epsfxsize = 7.5 cm  \epsffile{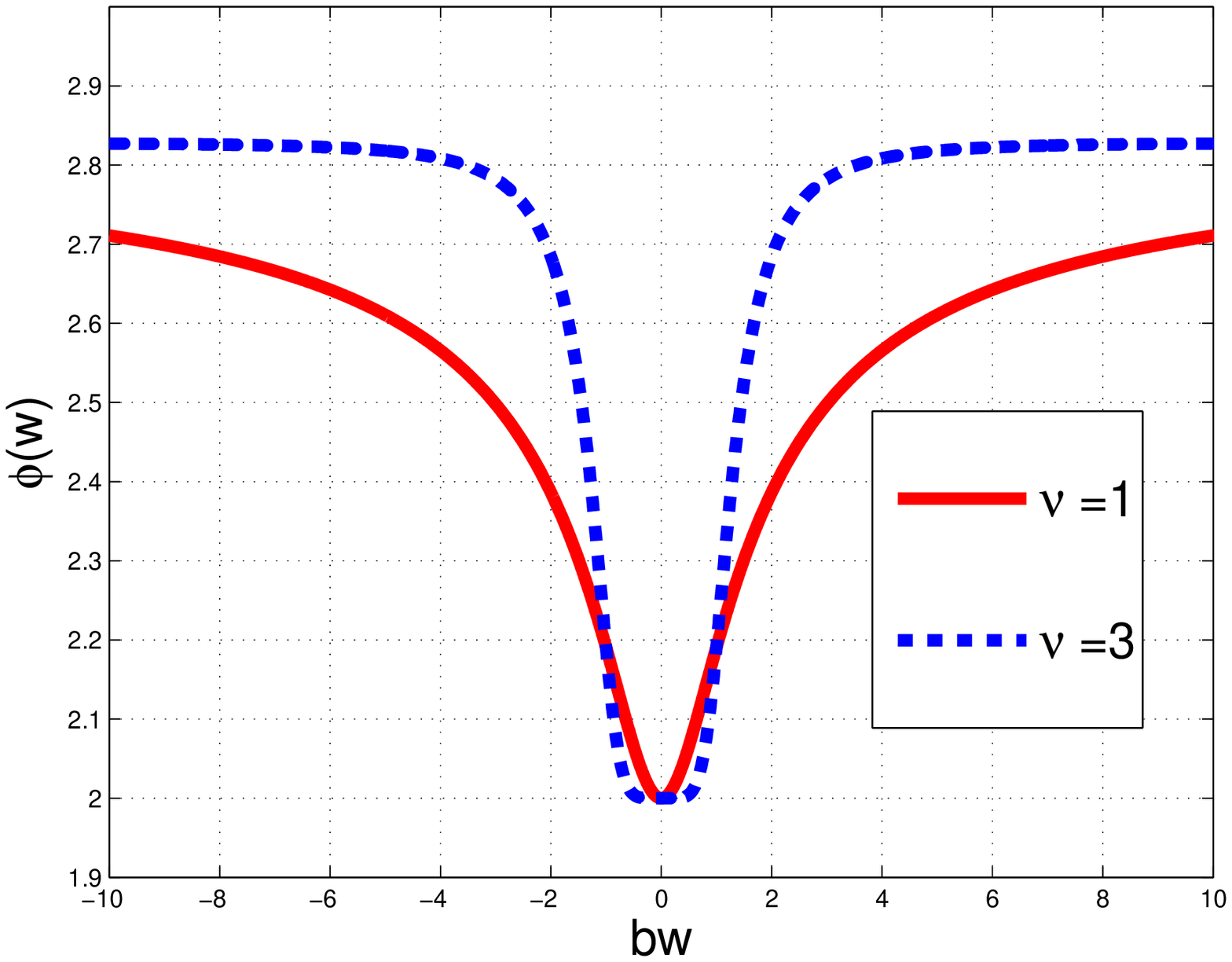}} &
      \hbox{\epsfxsize = 7.5 cm  \epsffile{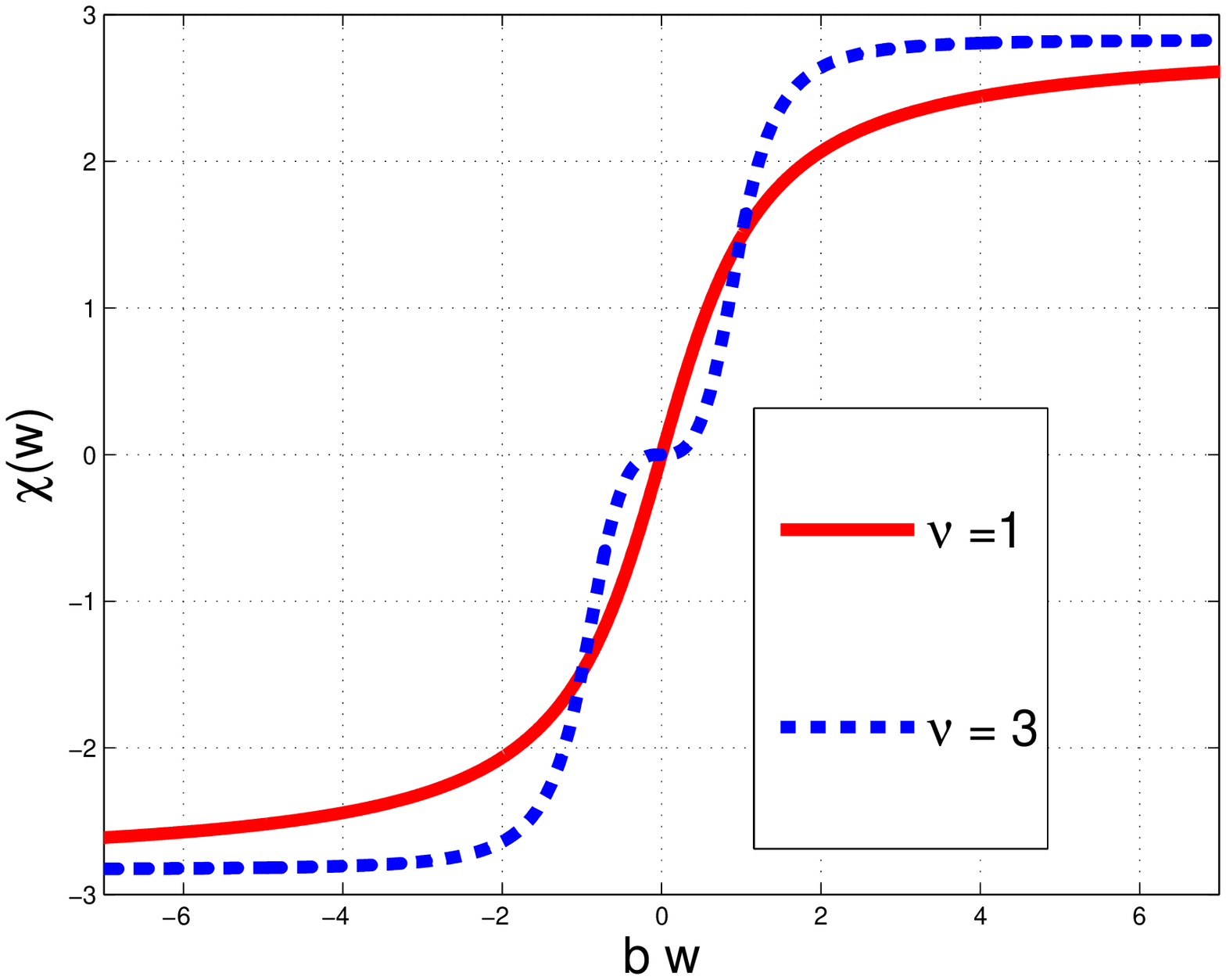}}\\
      \hline
\end{tabular}
\end{center}
\caption[a]{The trapping bag solutions arising in the case 
of odd values of $\nu$ are illustrated for few typical cases 
and by choosing, as in Fig. \ref{F1}, $v=1$. The full lines refer to the case 
$\nu=1$ while the dashes lines to the case $\nu = 3$. It should be appreciated
that the topological (plot at the right) and the non-topological (plot at the left) 
profiles are both simultaneously present for each odd value of $\nu$.}
\label{F2}
\end{figure}
In Fig. \ref{F2} the behaviour of $\phi(w)$ (left panel) and of $\chi(w)$ 
(plot at the right) is illustrated as a function of the dimensionless 
bulk coordinate $x = b w$ for $\nu = 1$ and $\nu =3$. 
Even if $\phi$ and $\chi$ are solutions, for a given 
value of $\nu$, of the same system of equations, the properties 
of the obtained defects are different. In particular, for odd values of 
$\nu$, $\phi$ always describes a non-topological profile.
More specifically,  Eq. (\ref{phNT}) imply
\begin{equation}
\lim_{w \to \pm \infty} \phi(w) = 2\sqrt{2} v.
\label{limitPH}
\end{equation}
Note that in Fig. \ref{F2} (as well in all the other figures) the value 
of $v$ has been set to $1$ in natural gravitational units 
$2\kappa =1$.

For the same odd values of $\nu$, the corresponding solution for $\chi(w)$ 
describes instead a topological soliton. In this case (see Fig. \ref{F2}, right 
plot) the field profile interpolates between $-2^{3/2} v$ and $+ 2^{3/2} v$:
\begin{equation}
\lim_{w\to -\infty} \chi(w) = -2 \sqrt{2}v,\qquad \lim_{w\to +\infty} \chi(w) = +2 \sqrt{2}v.
\end{equation}
In the case $\nu  = 3$ the solution also exhibits a short plateau 
centered around $w =0$.
Larger (odd) values of $\nu$ lead to profiles which are 
qualitatively similar to the ones discussed in the cases of Fig. \ref{F2}.
It is therefore legitimate to conclude that, for odd values of $\nu$, 
the solution defined by Eqs. (\ref{phNT}), (\ref{chNT}) and 
(\ref{hcond2}) always describe a bound system of a 
topological soliton (the $\chi$ field) and of a non-topological defect
 (the $\phi$ field). Similar solutions also arise in $(1+1)$-dimensional 
 field theories with appropriate non-linear potentials, and, in that 
 context, they have been named "trapping bags" \cite{RAJ1,MONT,WEI} since, in our language, the 
 $\chi$ field is "trapped" in the "bag" provided 
 by the $\phi$ profile.

From Eq. (\ref{hcond1}) it follows that if $h(x)$ is a solution, also 
$-h(x)$ will be a solution. If $ h \to - h$, then, from Eqs. (\ref{phNT}) 
and (\ref{chNT}), the r\^ole of $\phi$ and $\chi$ will be interchanged.
In particular, in the case of odd $\nu$, $h\to -h$ would imply that 
the topological profile becomes non-topological and 
vice-versa.
\begin{figure}
\begin{center}
\begin{tabular}{|c|c|}
      \hline
      \hbox{\epsfxsize = 7.5 cm  \epsffile{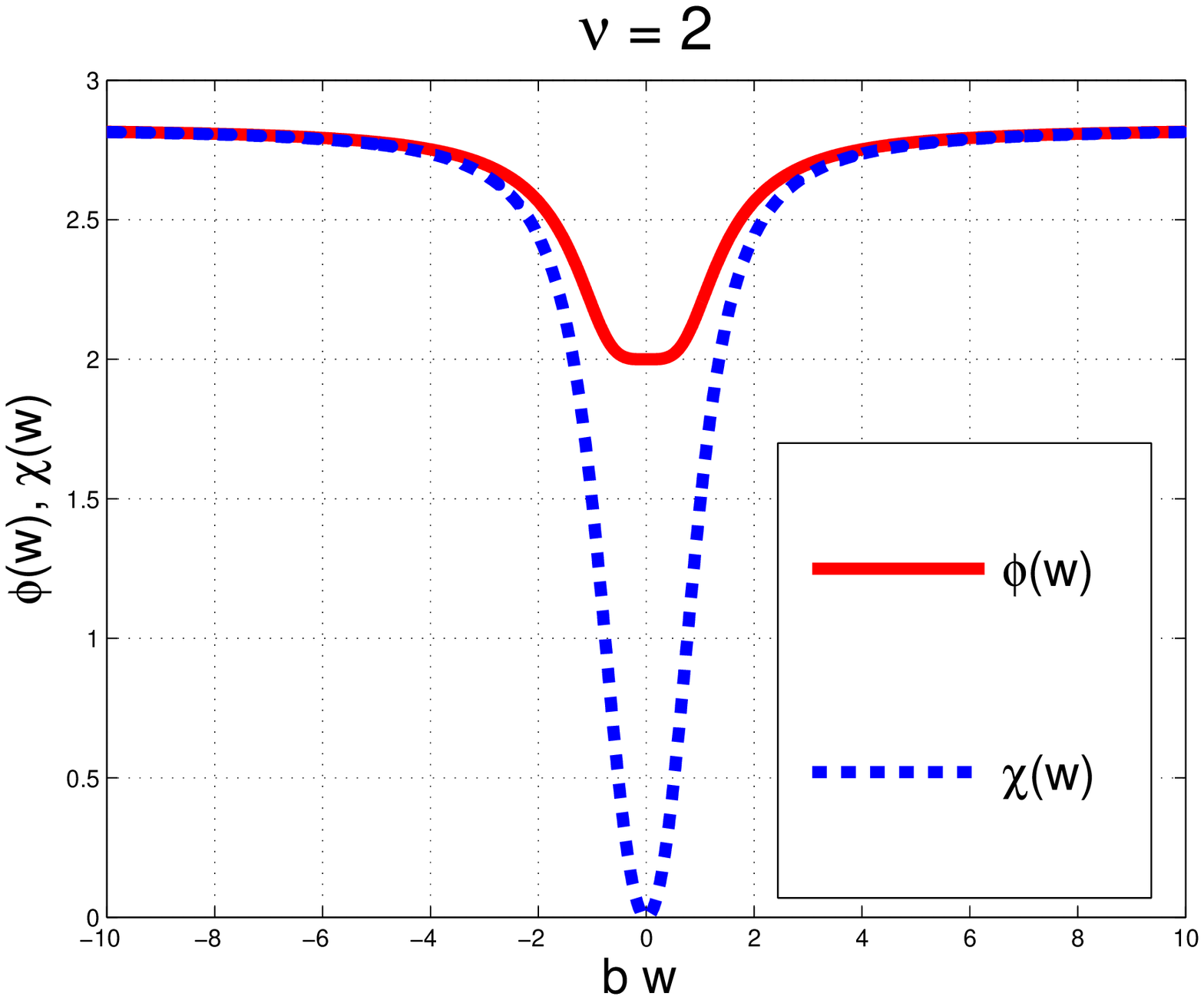}} &
      \hbox{\epsfxsize = 7.5 cm  \epsffile{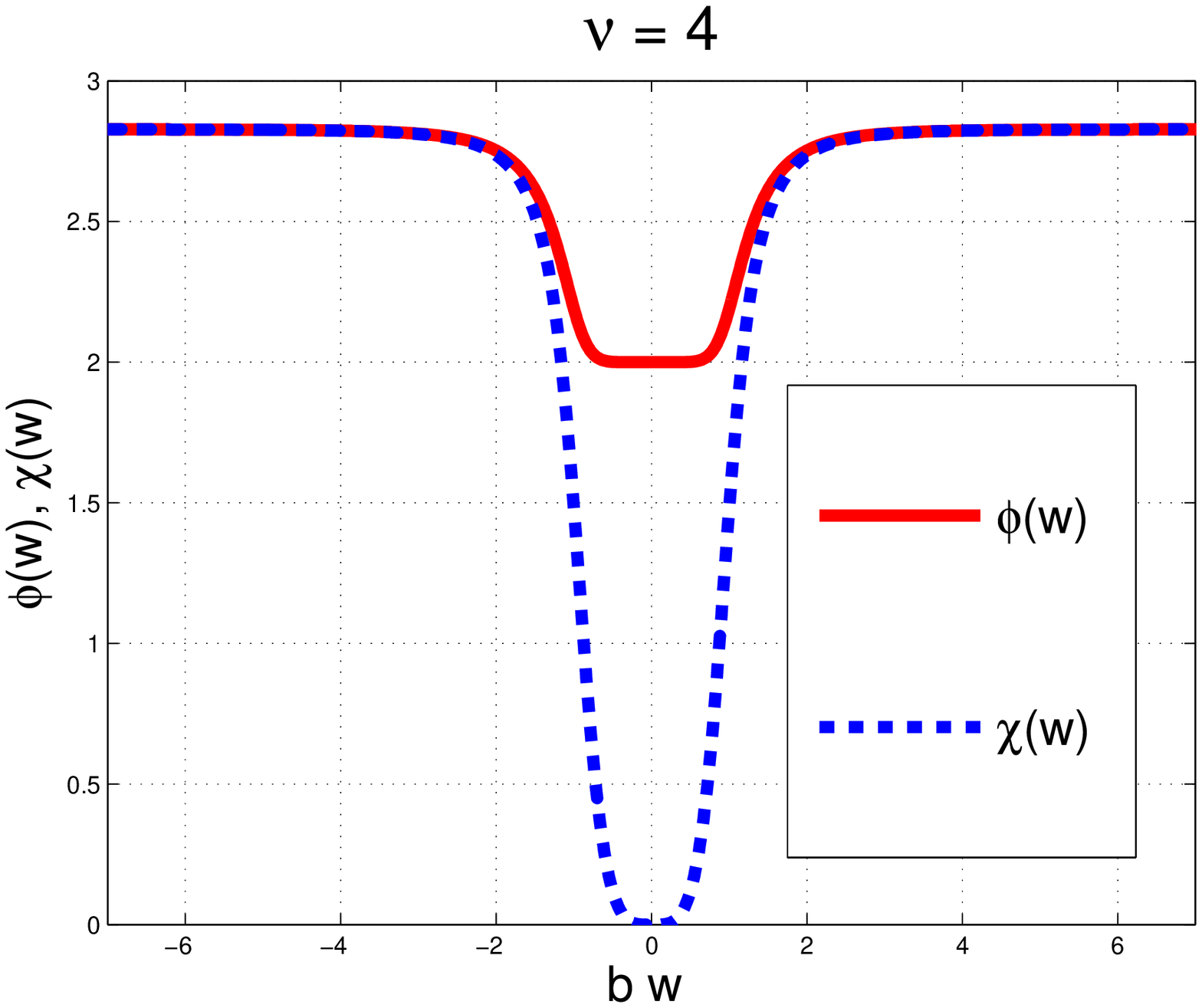}}\\
      \hline
\end{tabular}
\end{center}
\caption[a]{The bag-bag solutions arising in the case of even values of $\nu$
are illustrated for $\nu=2$ (left plot) and for $\nu =4$ (right plot). The value 
of $v$ is, as usual, set to $1$ in natural gravitational units. }
\label{F3}
\end{figure}
Let us now consider the case when $\nu$ is an even integer, i.e. 
$\nu = 2, 4, 6,...$.  In Fig. \ref{F3} the first two relevant values of $\nu$ 
are illustrated. In this situation it is not difficult 
to get convinced, by looking at Eqs. (\ref{phNT}), (\ref{chNT}) 
and (\ref{hcond2}) that $\phi(w)$ and $\chi(w)$ both describe 
non-topological profiles and that the overall solution is therefore 
a bound system of two bags, according to the terminology 
previously employed in the present section. This
statement can be explicitly verified since, in the case 
of even $\nu$, it holds that 
\begin{equation}
\lim_{w\to \pm \infty} \phi(w) = \lim_{w\to \pm \infty} \chi(w) = 2\sqrt{2}\, v.
\label{limiteven}
\end{equation}
It can be argued from Fig. \ref{F3} that as $\nu$ increases the width 
of the bag also increases. Moreover, both $\phi$ and $\chi$ 
have a global minimum for $w =0$. However, while $\chi(0) =0$
$\phi(0) = 2 \,v$. The configurations arising in the case of even values 
of $\nu$ illustrate the system formed of two bag profiles (i.e. bag-bag solutions).

\renewcommand{\theequation}{5.\arabic{equation}}
\section{Zero-modes for multidefects systems}
\setcounter{equation}{0}
In the following the techniques for the analysis of the zero modes in the 
case of gravitating multidefects will be presented.
The treatment followed in the present section generalizes 
the gauge-invariant approach to the fluctuations of single defects 
developed in \cite{MG1,MG2}. The main differences of the discussion
resides in the analysis of the scalar modes. In the case 
of a single gravitating defect in five-dimensional warped geometries it was shown that 
\begin{itemize}
\item{} the scalar modes of the configurations can be found 
by solving a single second-order (linear) differential equation 
for a gauge-invariant variable which is the combination of the fluctuations
of the geometry and of the fluctuation of the defect;
\item{} this gauge-invariant degree of freedom is indeed the canonical 
normal mode of the system, as it was demonstrated \cite{MG3} by diagonalizing 
the full second-order action for the scalar modes of the geometry 
coupled with the fluctuations of the defect;
\item{} the zero-mode (i.e. the lowest mass eigenstate) was shown 
not to be normalizable.
\end{itemize}
The introduction of a further scalar degree of freedom entails necessarily 
the need of a second gauge-invariant variable. While the generalization 
of the tensor and vector problem will be swiftly 
discussed just for completeness, the treatment of the scalar modes 
is qualitatively different with respect to single-defect configurations.

\subsection{General considerations}

Denoting by $\delta \phi$ and $\delta \chi$ the first-order perturbations of $\phi$ and $\chi$, the scalar degrees of freedom will be given by their 
background values supplemented by the appropriate fluctuations:
\begin{equation}
\phi \to \phi(w) + \delta \phi(x^{\mu}, w), \qquad \chi \to \phi(w) + \delta \chi(x^{\mu}, w).
\label{shift}
\end{equation}
By taking the trace of Eq. (\ref{EH1}) we get 
\begin{equation}
R_{AB} = \kappa \biggl(T_{AB} - \frac{T}{3}  G_{AB} \biggr), \qquad T_{A}^{A} = T,
\label{traced}
\end{equation}
which is particularly useful in perturbation theory since it 
gives automatically all the relevant equations avoiding the calculation 
of the first-order fluctuation of the Ricci scalar.
Therefore, by taking the first-order fluctuation of Eq. (\ref{traced}) 
and by recalling Eq. (\ref{TAB}) we will have 
\begin{eqnarray}
\delta R_{AB} &=& \kappa[ \partial_{A} \delta\phi \partial_{B} \phi + \partial_{A} \phi \partial_{B} \delta\phi
+ \partial_{A} \delta\chi \partial_{B} \chi + \partial_{A} \chi \partial_{B} \delta \chi ]
\nonumber\\
&-& \frac{2 \kappa}{3} W(\phi,\chi) \delta G_{A B} - \frac{2 \kappa}{3} \overline{G}_{A B} \biggl( 
\frac{\partial W}{\partial\phi} \delta \phi + \frac{\partial W}{\partial \chi} \delta \chi\biggr),
\label{pertEH}
\end{eqnarray} 
where $\overline{G}_{AB}$ denotes the background value of the metric tensor while $\delta G_{A B}$ and 
$\delta R_{AB}$ denote, respectively, the fluctuation of the five-dimensional metric and 
the fluctuation of the five-dimensional Ricci tensor. 

With similar notations, the first-order version of Eqs. (\ref{EH2}) and (\ref{EH3}) can be written by recalling 
that, when acting on a scalar degree of freedom, $\nabla_{A}\nabla_{B} = \partial_{A} \partial_{B} - \Gamma_{AB}^{C} \partial_{C}$ where $\Gamma_{A B}^{C}$ is the five-dimensional Christoffel connection. With 
this specification in mind we will have that:
\begin{eqnarray}
&&\delta G^{AB} [ \partial_{A} \partial_{B} \phi - \overline{\Gamma}_{A B}^{C} \partial_{C} \phi] + 
\overline{G}^{A B} [ \partial_{A} \partial_{B} \delta \phi - \delta \Gamma_{A B}^{C} \partial_{C} \phi - 
\overline{\Gamma}_{AB}^{C} \partial_{C} \delta \phi] 
\nonumber\\
&&+ 
\frac{\partial^2 W}{\partial \phi^2} \delta \phi + \frac{\partial^2 W}{\partial \phi \partial \chi} \delta \chi =0,
\label{phpert}\\
&&\delta G^{AB} [ \partial_{A} \partial_{B} \chi - \overline{\Gamma}_{A B}^{C} \partial_{C} \chi] + 
\overline{G}^{A B} [ \partial_{A} \partial_{B} \delta \chi - \delta \Gamma_{A B}^{C} \partial_{C} \chi - 
\overline{\Gamma}_{AB}^{C} \partial_{C} \delta \chi] 
\nonumber\\
&&+ 
\frac{\partial^2 W}{\partial \chi^2} \delta \chi + \frac{\partial^2 W}{\partial \phi \partial \chi} \delta \phi =0.
\label{chpert}
\end{eqnarray}
In Eqs. (\ref{phpert}) and (\ref{chpert}), $\overline{\Gamma}_{AB}^{C}$ and $\delta \Gamma_{A B}^{C}$ 
denote, respectively, the background Christoffel and their first-order fluctuations.

As usual \cite{MG1}, the fifteen degrees of freedom constituting $\delta G_{AB}$ in five dimensions\footnote{While it is 
rather simple to discuss the same decomposition in the case of $d$ transverse dimensions \cite{IMP}, we will be mainly concerned here with the case 
$d =3$.} can be 
separated, to first-order, into tensor, vector and scalar modes. 
In particular, this decomposition amounts to parametrize the various perturbed 
components of the metric tensor as 
\begin{eqnarray}
&& \delta G_{\mu\nu} = 2 a^2 h_{\mu\nu} + a^2 (\partial_{\mu} f_{\nu} +\partial_{\nu} f_{\mu}) + 
2 a^2 \eta_{\mu\nu} \psi  + 2 a^2 \partial_{\mu}\partial_{\nu} E,
\label{Gmunu}\\
&& \delta G_{\mu w} = a^2 (D_{\mu} + \partial_{\mu} C),
\label{Gmuw}\\
&& \delta G_{ww} = 2 a^2 \xi, 
\label{Gww}
\end{eqnarray}
where the Lorentz indices run over the four space-time dimensions. 
The tensor modes, i.e. $h_{\mu\nu}$, are, by definition, traceless and 
divergenceless, i.e. $\partial_{\mu} h^{\mu}_{\nu} = h_{\mu}^{\mu}$. 
Furthermore, it can be shown that, under infinitesimal coordinate 
transformations $h_{\mu\nu}$ is invariant. The invariance 
of a given fluctuations with respect to infinitesimal coordinate transformations 
will be referred to as gauge-invariance since the group of infinitesimal 
diffeomorphisms is, effectively, the gauge group of gravitation. 

The vector modes are parametrized by two divergenceless vectors 
i.e. $f_{\mu}$ and $D_{\mu}$ subjected to the conditions 
 $\partial_{\mu} D^{\mu}=0$ and $\partial_{\mu} f^{\mu}=0$. Overall there are 
 six  degrees of freedom.  
 
 Finally, the scalar modes are parametrized in terms of the four Lorentz scalars
 $\psi$, $\xi$, $E$ and $C$ whose evolution equations are coupled 
 with the fluctuations of the multidefect, i.e. $\delta \phi$ and $\delta\chi$.
 
 \subsection{Tensor zero-mode}
 The evolution equation 
 of the tensor modes is given by 
 \begin{equation}
 h_{\mu\nu}'' + 3 {\mathcal H} h_{\mu\nu}' - \Box h_{\mu\nu} =0,
 \label{tensoreq1}
 \end{equation}
 where $\Box = \eta^{\alpha\beta}\partial_{\alpha}\partial_{\beta}$. The zero mode of the tensor 
 fluctuations is localized in all the solutions presented in the previous sections
 owing to the quasi-$\mathrm{AdS_{5}}$ nature of the geometry. Defining, in fact, the rescaled variable
 $\mu_{\mu\nu} = a^{3/2} h_{\mu\nu}$,  Eq. (\ref{tensoreq1}) becomes:
 \begin{equation}
 \mu_{\mu\nu}'' - \Box \mu_{\mu\nu} - \frac{(a^{3/2})''}{a^{3/2}} \mu_{\mu\nu} =0.
 \label{tensoreq2}
 \end{equation}
 Note that $\mu_{\mu\nu}$ is the variable which appears canonically normalized 
 in the (second-order) action. Neglecting the Laplacian, the solution of Eq. (\ref{tensoreq2}) which is normalizable is given by $\mu_{\mu\nu} = c_{\mu\nu} a^{3/2}$
 where $c_{\mu\nu}$ does not depend on $w$ and $\mu_{0} (w) = a^{3/2}(w)$. Thus, the tensor zero mode is normalizable provided the integral
 \begin{equation}
 \int_{-\infty}^{+\infty} |\mu_{0}(w)|^2\, dw= \int_{-\infty}^{+\infty} a^3(w)\, dw 
 \label{normtens}
 \end{equation}
 is finite. This normalizability condition is satisfied by all the explicit forms of the 
 warp factors discussed so far in the present paper and, in particular 
 by Eqs. (\ref{ADS5}) and (\ref{wfnu}).
 In particular, it turns out that 
 \begin{eqnarray}
&& \int_{-\infty}^{\infty} \frac{dw}{[ 1 + (b w)^2]^{3/2}} = \frac{2}{b}
 \nonumber\\
 && \int_{-\infty}^{\infty} \frac{dw}{[ 1 + (b w)^{2\nu}]^{\frac{3}{2\nu}}} =
 \frac{2}{b}\frac{\Gamma\biggl(1 + \frac{1}{2\nu}\biggr) \Gamma\biggl(\frac{1}{\nu}\biggr)}{\Gamma\biggl(\frac{3}{2\nu}\biggr)},
 \label{INT}
\end{eqnarray}
where the result of the second integral in (\ref{INT}) is obtained under the assumption that $\nu$ is a natural number, i.e. the same assumption 
for which the solutions reported in sec. 4 are defined.
The finiteness of the integral 
 appearing in Eq. (\ref{normtens}) is also required by the finiteness 
 of the four-dimensional Planck mass
 \begin{equation}
 M_{\rm P}^2 = M_{5}^3 \int_{-\infty}^{+\infty} a^3(w) dw.
 \label{4DPL}
 \end{equation}
 
 \subsection{Vector zero-mode}
 For infinitesimal coordinate transformations 
 \begin{equation}
 x^{A} \to \tilde{x}^{A} = x^{A} +\epsilon^{A},\qquad \epsilon_{A} = a^2(w)( \epsilon_{\mu}, - \epsilon_{w}),
 \label{trans}
 \end{equation}
the scalar and vector modes of the geometry are transformed. In particular, 
by isolating the divergence-less part of $\epsilon_{\mu}$ as 
\begin{equation}
\epsilon_{\mu} = \partial_{\mu} \epsilon + \zeta_{\mu}, \qquad 
\partial_{\mu} \zeta^{\mu} =0,
\label{DIV}
\end{equation}
the gauge variation of the vector modes is 
\begin{eqnarray}
&& f_{\mu} \to \tilde{f}_{\mu} = f_{\mu} - \zeta_{\mu},
\nonumber\\
&& D_{\mu}\to \tilde{D}_{\mu} = D_{\mu} - \zeta_{\mu}'.
\label{vecvar}
\end{eqnarray}
From Eq. (\ref{vecvar}) it is immediate to ascertain 
the gauge-invariant vector mode which is given by 
\begin{equation}
V_{\mu} = D_{\mu} - f'_{\mu}.
\end{equation}
From the explicit form of Eq. (\ref{pertEH}) it is possible 
to obtain, after some algebra, that the 
zero-mode of $V_{\mu}$ obeys the following 
equation
\begin{equation}
{\mathcal V}_{\mu}' + \frac{3}{2} {\mathcal H} {\mathcal V}_{\mu} =0, \qquad 
{\mathcal V}_{\mu} = a^{3/2} V_{\mu}.
\label{veceq}
\end{equation}
From Eq. (\ref{veceq}) the normalizability of the vector zero-mode 
implies the convergence 
of the following integral 
\begin{equation}
\int_{-\infty}^{\infty} \frac{d w}{a^3(w)}. 
\label{VECnorm}
\end{equation}
It is clear that the normalizability condition implied by Eq. (\ref{VECnorm}) 
is, somehow, opposite to the one arising from Eqs. (\ref{normtens}) and 
(\ref{4DPL}). The conclusion is, therefore, exactly the same one would get 
in the case of single defects: if the graviton is normalizable, then the Planck mass is finite and the graviphoton is not normalizable. 
It should be remarked that this conclusion is typical of five-dimensional warped geometries. However, in even 
higher dimensions the situation may change. In particular, six-dimensional examples \cite{max61,max62} seem
to suggest that some vector degrees of freedom may indeed be localized (see also \cite{max63} for warped 
cosmological backgrounds in six dimensions).

\subsection{Scalar zero-modes}

While the discussion of the tensor and vector modes of the geometry 
mirrors completely the case of single defects, the scalar mode 
present relevant qualitative  differences. 
 
 The scalar modes of the geometry are not gauge-invariant 
 and are, in some sense, the most relevant ones since they are coupled 
 with the fluctuations of the multidefect configurations. 
 More specifically, for the gauge transformation written in Eqs. (\ref{trans}) 
 and (\ref{DIV}) the scalar fluctuations of the geometry change as 
 \begin{eqnarray}
 && \psi \to \tilde{\psi} = \psi - {\mathcal H} \epsilon_{w},
 \label{psig}\\
 && \xi\to \tilde{\xi} = \xi + {\mathcal H} \epsilon_{w} + \epsilon_{w}',
 \label{xig}\\
 && C \to \tilde{C} = C - \epsilon' + \epsilon_{w},
 \label{Cg}\\
 && E\to \tilde{E} = E - \epsilon.
 \label{eg}
 \end{eqnarray}
 To these transformations, the gauge variations of the fluctuations 
 of the defects must be added and they are 
 \begin{eqnarray}
 && \delta \phi \to \tilde{\delta\phi} = \delta\phi - \phi' \epsilon_{w},
 \label{dphig}\\
 && \delta\chi \to \tilde{\delta\chi} = \delta\chi - \chi' \epsilon_{w}.
 \label{dchig}
 \end{eqnarray}
 From Eqs. (\ref{psig})--(\ref{dchig}) it is clear that while gauge transformations 
 involving $\zeta_{\mu}$ preserve the vector nature 
 of the fluctuation, the transformations involving $\epsilon$ and $\epsilon_{w}$ 
 preserve the scalar nature of the fluctuation.
 
 It is possible, in the context of the scalar modes, to define a set 
 of gauge-invariant variables which are, in their simplest 
 incarnation:
 \begin{eqnarray}
 && \Psi = \psi - {\mathcal H} (E' - C),
 \label{PSI}\\
 && \Xi = \xi - \frac{1}{a} [ a(C - E')]' ,
 \label{XI}\\
 && X = \delta\chi + \chi' ( C- E'),
 \label{X}\\
 && \Phi = \delta \phi + \phi' (C- E').
 \label{PHI}
 \end{eqnarray}
In terms of the fluctuations defined 
in Eqs. (\ref{PSI})--(\ref{PHI}) the evolution equations of the scalar problem 
can be written in fully gauge-invariant terms. In particular, from the diagonal 
components of Eq. (\ref{pertEH}), i.e. the components 
$(\mu\mu)$ and $(ww)$, the following pair of equations can be obtained 
\begin{eqnarray}
&& \Box \Xi + 4 \Psi'' + 4 {\mathcal H} (\Psi' + \Xi') - 
\frac{4}{3} \kappa W a^2 \Xi + 2 \kappa [\phi' \Phi' + \chi' X']
\nonumber\\
&& + 
\frac{2}{3} \kappa a^2 \biggl[ \frac{\partial W}{\partial \phi} \Phi 
+ \frac{\partial W}{\partial \chi}X\biggr] =0
\label{mumu}\\
&& \Psi'' + 7 {\mathcal H} \Psi' + {\mathcal H} \Xi'  - \Box \Psi + 
2 ( {\mathcal H}' + 3 {\mathcal H}^2) \Xi + 
\frac{2}{3} \kappa a^2 \biggl[ \frac{\partial W}{\partial \phi} \Phi + 
\frac{\partial W}{\partial \chi} X\biggr] =0.
\label{ww}
\end{eqnarray}
From the off-diagonal elements of Eq. (\ref{pertEH}), i.e. $(\mu\neq \nu)$ and 
($\mu w$) we do get the following 
two conditions 
\begin{eqnarray}
&& \partial_{\mu}\partial_{\nu}[ \Xi - 2 \Psi] =0
\label{muneqnu}\\
&& 3 (\Psi' + {\mathcal H} \Xi) + \kappa (\phi' \Phi + \chi' X) =0.
\label{muw}
\end{eqnarray}
Finally, from Eqs (\ref{phpert}) and (\ref{chpert}) we get the evolution equations for 
$\Phi$ and $X$, i.e. the gauge-invvariant fluctuations of the defects:
\begin{eqnarray}
&& \Phi'' + 3 {\mathcal H} \Phi' - \Box \Phi - \frac{\partial^2 W}{\partial\varphi^2} a^2 \Phi - 
\frac{\partial^2 W}{\partial \phi \partial \chi} a^2 X 
\nonumber\\
&&+ \phi'[ 4 \Psi' + \Xi'] + 2
(\phi'' + 3 {\mathcal H} \phi') \Xi =0,
\label{PHIeq}\\
&& X'' + 3 {\mathcal H} X' - \Box X - \frac{\partial^2 W}{\partial\varphi^2} a^2 X - \frac{\partial^2 W}{\partial \phi \partial \chi} a^2 \Phi 
\nonumber\\
&&+ \chi'[ 4 \Psi' + \Xi'] + 2
(\chi'' + 3 {\mathcal H} \chi') \Xi =0.
\label{Xeq}
\end{eqnarray}

By subtracting Eqs. (\ref{mumu}) and (\ref{ww}) and by using the constraint 
$\Xi = 2 \Psi$ the following equation can be obtained 
\begin{equation}
\Psi'' + {\mathcal H} \Psi' + \Box \Psi = - \frac{2 \kappa}{3}[ \phi' \Phi' + \chi' X'].
\label{intpsi2}
\end{equation}
Moreover, using, again, the condition stemming from Eq. (\ref{muneqnu})
it is easy to reduce Eqs. (\ref{PHIeq}) and (\ref{Xeq}) to the following 
(more tractable) form:
\begin{eqnarray}
&& \Phi'' + 3 {\mathcal H} \Phi' - \Box \Phi - \frac{\partial^2 W}{\partial\varphi^2} a^2 \Phi - 
\frac{\partial^2 W}{\partial \phi \partial \chi} a^2 X + 6 \Psi' \phi'  
+ 4 (\phi'' + 3 {\mathcal H} \phi') \Psi =0,
\label{Phieqmod}\\
&& X'' + 3 {\mathcal H} X' - \Box X - \frac{\partial^2 W}{\partial\varphi^2} a^2 X - \frac{\partial^2 W}{\partial \phi \partial \chi} a^2 \Phi 
+ 6 \chi' \Psi' + 4 (\chi'' + 3 {\mathcal H} \chi') \Psi =0.
\label{Chieqmod}
\end{eqnarray}
In the presence of a single defect, the system describing the scalar modes of the geometry and of the sources can be put in a diagonal form by 
introducing the appropriately normal mode of the scalar system which is 
linear combination, (with background-dependent coefficients) of the fluctuation of the geometry and of the fluctuations of the defect \cite{MG3}.
 If two (or more) defects 
are simultaneously present this strategy can be generalized with important 
 qualitative differences. 
Let us therefore define the following pair of variables:
\begin{eqnarray}
&&{\mathcal G} = a^{3/2} \Phi - z_{\phi} \Psi,
\label{G}\\
&& {\mathcal F} = a^{3/2} X - z_{\chi} \Psi,
\label{F}
\end{eqnarray}
where 
\begin{equation}
z_{\phi} = \frac{ a^{3/2} \phi'}{{\mathcal H}},\qquad 
z_{\chi} = \frac{ a^{3/2} \chi'}{{\mathcal H}}.
\label{z}
\end{equation}
 Equations (\ref{G}) and (\ref{F}) can be used to eliminate $\Phi$ and $X$ from Eqs. (\ref{Phieqmod}) and (\ref{Chieqmod}). The same procedure can be 
 adopted in Eq. (\ref{intpsi2}): the dependence upon $\Phi$ and $X$ appearing 
 at the right hand side off Eq. (\ref{intpsi2}) can be eliminated in favour of ${\mathcal G}$ and ${\mathcal F}$. The obtained equation can be 
 used to eliminate the dependence on $\Psi$ arising in Eqs. (\ref{Phieqmod}) 
 and (\ref{Chieqmod}). The net result of this procedure is that the evolution equations for ${\mathcal G}$ and ${\mathcal F}$ is given by
 \begin{eqnarray}
&& {\mathcal G}'' - \Box {\mathcal G} -
 {\mathcal M}_{{\mathcal G}{\mathcal G}} {\mathcal G} 
 -  {\mathcal M}_{{\mathcal G}{\mathcal F}} {\mathcal F} =0,
 \label{Geq}\\
 && {\mathcal F}'' - \Box {\mathcal F} -
 {\mathcal M}_{{\mathcal F}{\mathcal F}} {\mathcal F} 
 -  {\mathcal M}_{{\mathcal F}{\mathcal G}} {\mathcal G} =0,
 \label{Feq}
 \end{eqnarray}
 where 
 \begin{eqnarray}
 &&{\mathcal M}_{{\mathcal G}{\mathcal G}}= \frac{(a^{3/2})''}{a^{3/2}}  + 
 a^2 \frac{\partial^2 W}{\partial \phi^2} - \frac{2 \kappa}{3 {\cal H}^2} 
 {\mathcal H}' {\phi'}^2  + \frac{4}{3} \kappa \frac{\phi' \phi''}{{\mathcal H}} + 2 \kappa
 {\phi'}^2,
\label{Mgg}\\
&&{\mathcal M}_{{\mathcal F}{\mathcal F}}=
\frac{(a^{3/2})''}{a^{3/2}}  + 
 a^2 \frac{\partial^2 W}{\partial \chi^2} - \frac{2 \kappa}{3 {\cal H}^2} 
 {\mathcal H}' {\chi'}^2  + \frac{4}{3} \kappa \frac{\chi' \chi''}{{\mathcal H}} + 2 \kappa
 {\chi'}^2,
\label{Mff}\\
&& {\mathcal M}_{{\mathcal F}{\mathcal G}}= 
{\mathcal M}_{{\mathcal G}{\mathcal F}} = \frac{\partial^2 W}{\partial \phi \partial\chi}a^2
- \frac{2 \kappa {\mathcal H}'}{3 {\mathcal H}^2} \phi' \chi' + 
\frac{2\kappa}{3 {\mathcal H}} (\phi' \chi'' + \chi' \phi'') + 2 \kappa \phi' \chi'.
\label{Mfg}
\end{eqnarray}
In the limit $W(\phi,\chi) \to W(\phi)$ and $\chi'\to 0$, ${\mathcal M}_{{\mathcal G}{\mathcal F}}\to 0$ and ${\mathcal M}_{{\mathcal G}{\mathcal G}} \to z_{\phi}''/z_{\phi}$.
Thus, the result of the single defect system is recovered.

Defining 
\begin{equation}
{\mathcal L} = \pmatrix{ {\mathcal G}\cr
{\mathcal F}}, \qquad {\mathcal M} =  \pmatrix{{\cal M}_{{\mathcal G}{\mathcal G}} & {\cal M}_{{\mathcal G}{\mathcal F}}\cr
 {\mathcal M}_{{\mathcal G}{\mathcal F}}& {\mathcal M}_{{\mathcal F}, {\mathcal F}}},
 \label{matrix}
 \end{equation}
Eqs. (\ref{Geq}) and (\ref{Feq}) can be simply written in matrix notation
\begin{equation}
{\mathcal L}'' - \Box {\mathcal L} - {\mathcal M} {\mathcal L}=0.
\label{matrixeq}
\end{equation}

It is worth noticing that the derivation presented here 
is fully gauge-invariant. Since ${\mathcal G}$ and ${\mathcal F}$ 
are gauge-invariant their evolution equation is the same in any 
specific gauge. It is therefore useful to cross-check the gauge-invariant 
derivation with a gauge-dependent discussion. A particularly useful 
coordinate system, already exploited in \cite{MG2} is the 
off-diagonal gauge where $E=0$ and $\psi = 0$. The 
gauge-invariant variables become, in this gauge, 
\begin{eqnarray}
&& \Psi = {\mathcal H} C,\qquad \Xi = \xi - {\mathcal H} C - C',
\label{A1}\\
&& \Phi = \delta \phi + \phi' C,\qquad X = \delta \chi + \chi' C.
\label{A2}
\end{eqnarray}
Inserting Eqs. (\ref{A1}) and (\ref{A2}) into Eqs. (\ref{G}) and (\ref{F}) 
it is also clear that, in the off-diagonal gauge, 
${\mathcal G} = a^{3/2} \delta\phi$ and ${\mathcal F} = a^{3/2} \delta \chi$.
Therefore, using the evolution equations written explicitly in the off-diagonal gauge, it is possible to recover exactly Eqs. (\ref{Geq}) and (\ref{Feq}).
The essential steps of this exercise are summarized in the Appendix.

Let us now consider, as a useful application, the determination 
of the localization properties of the scalar zero-modes in the case 
of the solutions discussed in sec. 3 and parametrized according 
to Eq. (\ref{santis1}).
In the case of the kink-antikink system the various 
matrix elements appearing in Eq. (\ref{matrix}) can be 
computed in terms of $g$. The result of this algebraic calculation is  
\begin{eqnarray}
&&{\mathcal M}_{{\mathcal G}{\mathcal G}} = \frac{b^2}{4 g^2}
[g^2 (1- g^2) (76 g^2 + 42 g + 1) + 4 ( 11+ g)^2 ( 1- g)^3( 6g^2 + g + 1)],
\nonumber\\
&& {\mathcal M}_{{\mathcal F}{\mathcal G}} = {\mathcal M}_{{\mathcal G}{\mathcal F}} = \frac{b^2}{g^2} [ 7 g^2 ( 1 - g^2)^{3/2} + (1-g^2)^{5/2} (g+1)(2g -1) + 3 g ( 1-g^2)^2],
\nonumber\\
&& {\mathcal M}_{{\mathcal F}{\mathcal F}} = \frac{b^2}{4 g^2}[ g^2 ( 1 - g^2)
(76 g^2 - 42 g +1) + 4 ( 1- g)^2 (1+ g)^3( 6g^2 - g +1)],
\label{MAT1g}
\end{eqnarray}
Using the explicit expression of $g(x)$ derived in Eq. (\ref{gcond2}) 
the matrix elements of ${\mathcal M}$ can be expressed as a function 
of the rescaled bulk coordinate $x = bw$:
\begin{eqnarray}
{\mathcal M}_{{\mathcal G}{\mathcal G}} &=& \frac{b^2}{4 x^2  (x^2 +1)^{5/2}}
[x^2 \sqrt{x^2 +1} (77 x^2 + 42 x \sqrt{x^2 +1} + 1) 
\nonumber\\
&+ & 4 (\sqrt{x^2 + 1} - x)(7x^2 + x\sqrt{x^2 + 1} + 1)],
\nonumber\\
 {\mathcal M}_{{\mathcal F}{\mathcal G}} &=& {\mathcal M}_{{\mathcal G}{\mathcal F}} = \frac{b^2}{x^2 (x^2 +1)^{5/2}} [ x(x^2 + 1) (7x + 3) + (x^2 + x \sqrt{x^2+1} -1)],
\nonumber\\
{\mathcal M}_{{\mathcal F}{\mathcal F}}& =& \frac{b^2}{4 x^2 (x^2 +1)^{5/2}}[ 
x^2 \sqrt{x^2 +1} (77 x^2 - 42 x \sqrt{x^2 +1} +1) 
\nonumber\\
&+ &4 ( \sqrt{x^2 + 1} + x) ( 7x^2 - x\sqrt{x^2 + 1} + 1)].
\label{MAT2x}
\end{eqnarray}
To determine if the lowest mass eigenstates are normalizable or not 
it suffices to go see if Eqs. (\ref{Geq}) and (\ref{Feq}) admit 
normalizable solutions in the case $\Box {\mathcal G} = \Box{\mathcal F}=0$
when the coefficients ${\mathcal M}_{{\mathcal G}{\mathcal G}}$, 
${\mathcal M}_{{\mathcal F}{\mathcal G}}$ and ${\mathcal M}_{{\mathcal F}{\mathcal F}}$, are the ones determined in Eq. (\ref{MAT2x}).

The idea is therefore the following. Let us solve, asymptotically, the 
explicit form of Eqs. (\ref{Geq}) and (\ref{Feq}) in the limit 
$x\to -\infty$. By imposing initial condtions for $x\to -\infty$ the system 
can be integrated numerically across $x=0$. This procedure 
will give, as a function of different initial conditions, the lowest mass
eigenstates provided the obtained solution is regular and sufficiently 
convergent for large absolute values of the rescaled bulk coordinate.

In the limit $x\to -\infty$ it can be verified that Eq. (\ref{MAT2x}) 
gives
\begin{equation}
{\mathcal M}_{{\mathcal G}{\mathcal G}}\simeq \frac{119}{4} \frac{b^2}{x^2},\qquad
{\mathcal M}_{{\mathcal F}{\mathcal G}} ={\mathcal M}_{{\mathcal G}{\mathcal F}} \simeq \frac{7 b^2}{x^3},\qquad {\mathcal M}_{{\mathcal F}{\mathcal F}}
= \frac{35}{4} \frac{b^2}{x^2}.
\label{MAT3x}
\end{equation}
Therefore, for $x \to -\infty$, the solutions for ${\mathcal G}$ and ${\mathcal F}$ are simple power law, i.e.
\begin{eqnarray}
&& {\mathcal G}_{0}(x) \simeq |x|^{\gamma_{\pm}}, \qquad \gamma_{\pm}=\frac{1 \pm 2\sqrt{30}}{2},
\label{Gsol}\\
&& {\mathcal F}_{0}(x) \simeq |x|^{\delta_{\pm}}, \qquad \delta_{+}=\frac{7}{2},\qquad \delta_{-} = - \frac{5}{2}.
\label{Fsol}
\end{eqnarray}
Recall now that the zero mode is normalizable provided
\begin{equation}
\int_{-\infty}^{+\infty} |{\mathcal G}_{0}(x)|^2 \, dx,\qquad 
\int_{-\infty}^{+\infty} |{\mathcal F}_{0}(x)|^2 \, dx
\label{condfin}
\end{equation}
are both convergent. This requirement implies that the only 
asymptotic initial conditions (for $x \to -\infty$) compatible 
with the finiteness of the integrals appearing in Eq. (\ref{condfin}) 
are the ones parametrized in terms of $\gamma_{-}$ and $\delta_{-}$.
The numerical integration of the system can be performed by starting 
with a sufficiently negative $x_{\rm i}$, for instance  $x_{\rm i} = - 10^{3}$.
Thus, the initial conditions will be dictated by 
${\mathcal G}_{0}(x_{\rm i}) \simeq |x_{\rm i}|^{\gamma_{-}}$ and by ${\mathcal F}_{0}(x_{\rm i}) \simeq 
|x_{\rm i}|^{\delta_{-}}$. It turns
out, after explicit numerical integration,  that a singularity is always developed in the origin so that the obtained 
solution is not normalizable because of the behaviour near $x=0$ where 
the matrix elements of ${\mathcal M}$ diverge. We then conclude 
that the kink-antikink system does not admit a normalizable 
zero mode.
A similar discussion can be performed in the case of the solutions 
reported in section 4. Also in that case there scalar zero modes are not 
normalizable because of the behaviour of the solution in the origin.

\renewcommand{\theequation}{6.\arabic{equation}}
\section{Concluding remarks}
\setcounter{equation}{0}
The possibility of gravitating multidefects  from higher-dimensional warped 
geometries has been scrutinized. It has been shown 
that, indeed, it is possible to find systems where 
qualitatively different profiles arise simultaneously. 
After devising a general method for the 
construction of multidefects solutions, specific examples, compatible 
with $\mathrm{AdS_{5}}$, have been presented. While the obtained
solutions were only illustrative, it can be argued, on general grounds, that the following configurations 
can explicitly constructed:
\begin{itemize}
\item{} kink-antikink systems both in the case 
of Einstein-Hilbert gravity and in the case of Gauss-Bonnet 
gravity;
\item{} configurations containing one topological soliton and 
a non-topological profile;
\item{} systems containing two bag-like profiles, both 
non-topological.
\end{itemize}
The second class of solutions listed above 
corresponds, indeed, to the so-called trapping bag solutions 
that may be used to model (static) confining 
configurations in $(1+1)$ dimensions. 
As in the $(1+1)$ dimensional case three-defects 
solutions may be constructed, it is justified to speculate that 
the present considerations 
may also be extended to the case of three (or even more) defects.

As far as the localization properties of the zero modes are concerned, 
it has been shown that while the tensor and vector modes exhibit
exactly the same features arising in the case of single 
gravitating defects, the scalar modes present qualitatively 
new features. In the case of gravitating multidefects, the fluctuations 
of the geometry are coupled with the fluctuations of all the profiles
of the system. A fully gauge-invariant formalism for the 
analysis of the localization properties of the fluctuations 
of various spin has then been developed by extending 
the treatment already exploited in the case of single 
gravitating defects.
\newpage
\begin{appendix}
\renewcommand{\theequation}{A.\arabic{equation}}
\setcounter{equation}{0}
\section{Off-diagonal gauge}
In this Appendix the decoupled set of evolution equations for the scalar 
modes of the geometry will be derived in the off-diagonal gauge. 
The reported exercise complements and corroborates the 
gauge-invariant derivation reported in sec. 5. 
The rationale for the previous statement is that since 
the canonical (scalar) normal modes defined in Eqs. (\ref{G}) and (\ref{F})
are gauge-invariant, they must also obey the same evolution equations 
(i.e. Eqs. (\ref{Geq}) and (\ref{Feq})) in any specific gauge. As 
observed in sec.5, in the off-diagonal gauge, i.e. the gauge 
where $\psi=E=0$, ${\mathcal G} = a^{3/2} \delta\phi$ and ${\mathcal F} =
a^{3/2} \delta\chi$. The exercise presented here will the be to 
derive Eqs. (\ref{Geq}) and (\ref{Feq}) by using directly the evolution equations 
written in the off-diagonal gauge.

The evolution equations in the off-diagonal gauge can be obtained by expressing the 
gauge-invariant quantities of Eqs. (\ref{PSI})--(\ref{PHI}) in the gauge 
$\psi =0$ and $E=0$.  The result of this procedure is given by Eqs. (\ref{A1}) 
and (\ref{A2}).
Using Eqs. (\ref{A1}) and (\ref{A2}) into Eqs. (\ref{mumu})--(\ref{ww}) 
we get 
\begin{eqnarray}
&& {\mathcal H} \xi' + 2 ({\mathcal H}' + 3 {\mathcal H}^2) \xi - 
{\mathcal H} \Box C + 
\frac{2}{3} \kappa a^2 \biggl[ \frac{\partial W}{\partial \phi}  \delta \phi + 
\frac{\partial W}{\partial \chi} \delta \chi\biggr] =0
\label{mumu1}\\
&& {\mathcal H} \Box C + \Box C' - 4 {\mathcal H} \xi' - \Box \xi  - 2 \kappa [ \phi \delta\phi' + \chi' \delta\chi' ] 
\nonumber\\
&&- \frac{4}{3} \kappa a^2 W \xi+ \frac{2}{3} \kappa a^2 \biggl( 
\frac{\partial W}{\partial \phi} \delta \phi + \frac{\partial W}{\partial \chi} \delta\chi \biggr),
\label{ww1}
\end{eqnarray}
Using Eqs. (\ref{A1}) and (\ref{A2}) into Eqs. (\ref{muneqnu})--(\ref{muw}) 
the following pair of conditions are obtained:
\begin{eqnarray}
&& \xi - C' - 3 {\mathcal H} C =0
\label{muneqnu1}\\
&& 3 {\mathcal H} \xi = - \kappa (\phi'\delta \phi + \chi' \delta \chi),
\label{muw1}
\end{eqnarray}
Finally, using Eqs. (\ref{A1}) and (\ref{A2}) into Eqs. (\ref{PHI}) and (\ref{X}) 
we do get:
\begin{eqnarray}
&& \delta\phi'' +3 {\mathcal H} \delta\phi' - \Box \delta\phi - 
\frac{\partial^2 W}{\partial \phi^2} a^2 \delta \phi - \frac{\partial^2 W}{\partial\phi\partial\chi} a^2 \delta\chi 
\nonumber\\
&&+ (\xi' - \Box C) \phi' + 2 (\phi'' + 3 {\mathcal H}\phi') \xi =0
\label{deltaphieq}\\
&& \delta\chi'' +3 {\mathcal H} \delta\chi' - \Box \delta\chi - 
\frac{\partial^2 W}{\partial \chi^2} a^2 \delta \chi - \frac{\partial^2 W}{\partial\phi\partial\chi} a^2 \delta\phi 
\nonumber\\
&&+ (\xi' - \Box C) \chi' + 2 (\chi'' + 3 {\mathcal H}\chi') \xi =0
\label{deltachieq}
\end{eqnarray}
As already mentioned, in the off-diagonal gauge the variables ${\mathcal G}$ and ${\mathcal F}$ become
\begin{equation}
{\mathcal G} = a^{3/2} \delta\phi,\qquad 
{\mathcal F} =  a^{3/2} \delta\chi.
\label{GFOD}
\end{equation}
Subtracting Eqs. (\ref{ww1}) from Eq. (\ref{mumu1})
and recalling Eq. (\ref{muw}) we do get 
\begin{eqnarray}
&&\Box C - \xi' = - \frac{2\kappa}{3 {\mathcal H}}[\phi' \delta\phi' + 
\chi' \delta\chi'] - 2 \xi',
\label{int1}\\
&& \xi = - \frac{\kappa}{3 {\mathcal H}}[ \phi' \delta\phi +\chi' \delta\chi].
\label{int2}
\end{eqnarray}
By inserting Eq. (\ref{GFOD}) into Eqs. (\ref{deltaphieq}) and (\ref{deltachieq}) 
$\delta \phi$ and $\delta\chi$ can be completely eliminated 
in favour of ${\mathcal G}$ and ${\mathcal F}$. In  Eqs. (\ref{deltaphieq}) and (\ref{deltachieq}) the only remaining 
pieces will be, respectively, the expressions proportional to 
$(\Box C - \xi')$ and to $\xi$:
\begin{eqnarray}
&& \frac{1}{a^{3/2}} \biggl\{ {\mathcal G}'' - \Box {\mathcal G} - \biggl[ \frac{(a^{3/2})''}{a^{3/2}} + \frac{\partial^2 W}{\partial\phi^2} a^2 \biggr] - \frac{\partial^2 W}{\partial \phi\partial\chi}{\mathcal F}\biggr\}
\nonumber\\
&&+ (\xi' - \Box C) \phi' + 2 (\phi'' + 3 {\mathcal H}\phi') \xi=0,
\label{intphi3}\\
&&\frac{1}{a^{3/2} }\biggl\{ {\mathcal F}'' - \Box {\mathcal F} - \biggl[ \frac{(a^{3/2})''}{a^{3/2}} + \frac{\partial^2 W}{\partial\chi^2} a^2 \biggr] - \frac{\partial^2 W}{\partial \chi\partial\phi}{\mathcal G}\biggr\}
\nonumber\\
&&+ (\xi' - \Box C) \chi' + 2 (\chi'' + 3 {\mathcal H}\chi') \xi=0.
\label{intchi3}
\end{eqnarray}
But using Eq. (\ref{GFOD}) into Eqs. 
(\ref{int1}) and (\ref{int2}) we do get 
\begin{eqnarray}
&& \Box C - \xi' = \frac{2\kappa}{3 {\mathcal H} a^{3/2}}\biggl[ 
\phi'' {\mathcal G} + \chi'' {\mathcal F} - \frac{{\mathcal H}'}{{\mathcal H}} ( \phi' 
{\mathcal G} + \chi' {\mathcal F})\biggr],
\label{int1ex}\\
&&\xi = - \frac{\kappa}{3 {\mathcal H} a^{3/2}}(\phi' {\mathcal G} + \chi' {\mathcal F}).
\label{int2ex}
\end{eqnarray}
Equations (\ref{int1ex}) and (\ref{int2ex}) show that $(\Box C - \xi')$ and to $\xi$ can be solely expressed in terms of ${\mathcal G}$, ${\mathcal F}$ and their 
first derivatives with respect to the bulk coordinate. This completes the derivation since inserting then Eqs. (\ref{int1ex}) and (\ref{int2ex}) into Eqs. 
(\ref{intphi3}) and (\ref{intchi3}),  Eqs. (\ref{Geq}) and (\ref{Feq})  
are readily obtained. 
\end{appendix}

\end{document}